# Information Technology in New Zealand: Review of Emerging Social Trends, Current Issues, and Policies

Emre Erturk, Derwyn Fail

**Abstract**

This paper discusses the general state of information technology in New Zealand society, current issues, and policies. It is a qualitative study that reviews recent scholarly articles, periodicals, and surveys in order to create an understanding of some of the information technology issues and trends in New Zealand. After reviewing previous research, it assesses the potential existence and nature of a 'digital divide' in New Zealand society whilst also evaluating possible strategic responses to the issue. New Zealand society has rapidly accepted emerging online trends as well as achieving an overall high level of Internet provision nationally. Through government policy and education, this small island nation has remained at the forefront of information technology and can be considered somewhat of an e-democracy. However, despite these positives, there is a risk of low-income communities being left behind as New Zealand society becomes increasingly dependent on IT in the workplace and in governmental administration.

## 1. Introduction

This paper analyses the current state of information technology in New Zealand society through the evaluation of various articles, surveys and research. The aim of this paper is to develop an accurate understanding of the current position New Zealand finds itself in with regards to information technology (IT), the effect IT has had on society, and the possible disadvantages caused by these developments. Although this is a qualitative review, it also draws on some of the previous quantitative research in order to provide informed conclusions and recommendations.

The advance of Information technology has significantly altered every aspect of modern New Zealand (NZ) society. The degree to which this has occurred is directly related to the penetration and saturation of information technology systems within the social, professional, political and government spheres in NZ society. This paper will discuss the effects, both positive and negative, that the IT transformation has had on different aspects of NZ society. The statistical and demographic profile of the NZ workforce has changed in response to the need for new skill sets, affecting even traditional 'blue collar' manual labour. The willingness of agricultural, manufacturing, retail, financial, and commercial sectors to embrace new technology, and the rapid growth of a professional middle class in response to this, have resulted in a workforce with more advanced technological proficiency, educated and trained in the new ways of working. However, any potentially damaging and isolating effects this transformation may have had on some social groups who, for various reasons, have 'missed out' on IT resources and skills, should be explored for any evidence of economic, political, or job market marginalisation and disenfranchisement.

## 2. Literature Review

In his article *The New Age of Business Mobile*, Kevin Kevany discusses the influence of advanced mobile devices and applications and their affects on New Zealand businesses and productivity. The article investigates exciting developments in mobile devices and their use in business, such as the support for "BYOD" in the workplace in New Zealand. 'Bring your own device' computing, or BYOD, is one of the major trends in the current business and computing environment. According to the IT research company 'Gartner', BYOD is the single most radical shift in the economics of client computing for business since computers began appearing in the workplace [17]. BYOD is the business strategy that allows employees to use their own devices to run enterprise applications, access data and carry out day-to-day business functions. A company's BYOD strategy is specifically targeted toward smart phones and tablets. There are many small-scale local examples that illustrate how IT and mobile technology have infiltrated the modern business environment in New Zealand. Rebecca McLeod's small New Zealand business 'Made4Baby' is a company that specializes in skincare products for newborn babies. She uses her smart phone throughout the day, making deliveries to keep stock records, processing orders and invoices, and updating the 'Made4Baby' Twitter and Facebook pages [8]. As opposed to going about her daily business and returning home to fill in the paperwork, Rebecca McLeod is able to carry out all these functions using mobile Internet. This important trend highlights the reliance of the NZ business sector on mobile technology.

A digital divide is an inequality between groups or countries in terms of access to, use of, or knowledge of information and communication technologies such as the Internet. The divide inside countries, i.e. a domestic digital divide which this paper will also discuss, refers to inequalities of Internet and Information technology use between individuals, households, businesses, and geographic areas at different socioeconomic and other demographic levels [13]. An interesting recent study concluded that a digital divide exists within the European Union [1]. There are countries with greater digital resources and higher use, and those at lower levels. Among countries with lower Internet adoption, there are those who are doing well in terms of business use but not as well for individual household use, or vice versa.

The quantitative study *The Digital Divide* was conducted by Statistics New Zealand in March 2004 to assess the main characteristics that influence Internet connectivity in New Zealand [15]. The data was drawn from the 2001 census with supporting data from the household economic survey. The summary of findings provides interesting information with regards to Internet usage within the average household and found that total household income had a direct influence on Internet usage. There was also declining access levels in older age groups, as well as a much higher usage in larger cities such as Wellington and Auckland compared to the smaller towns situated around New Zealand, both suggesting a potential digital divide emerging in New Zealand.

At that time, these statistics suggested a movement towards a digital divide in New Zealand particularly with rural areas, the elderly, and low-income families. Although the study was carried out in 2004, the conclusions it draws are interesting and worthy of investigation today. The results are still valuable because it details a global trend in characteristics that influence Internet connectivity, and more recent data has not shown any major differences to the findings of this paper. It is still evidenced that total household income has a direct effect on Internet. This is an important study because the information produced has influenced subsequent policy and responses to the digital divide in New Zealand. Cities will always have a higher Internet use then rural towns however there is an increasing effort to address this issue, the Rural Broadband initiative is an example of this with the aim to bring high speed Internet to 84% of rural customers by 2016.

*Addressing the 'Digital Divide' in Aotearoa New Zealand* was written for the Finnish Encyclopaedia of Digital Government by Marilyn Head and used to help draft the New Zealand government and communities Digital Strategy in 2005. It gives a general background of the physical, demographic, social, economic and political condition relevant to the use of IT in New Zealand. It discusses the widespread IT use in New Zealand society particularly with 88% of NZ businesses using computers for some form of E-business. Head identifies that there is innovative use of IT in the agriculture industry, particularly Fonterra's rural networking program aimed at linking dairy farmers all over New Zealand to provide a two way information flow between farms and Fonterra [5].

Various IT initiatives in New Zealand prior to 2004 include the E-government initiative aimed at the delivery of all government processes through the online channel by 2010. Project PROBE was another such initiative aimed at delivering high speed Internet to all schools and provincial communities on a region by region basis, and was highly successful at a cost of $ 48 million [5]. E-Education strategies such as the Digital Opportunities initiative was launched in 2001, with the aim of using IT to improve educational achievement particularly in mathematics, science and technology [5]. Other initiatives aimed at ensuring all aspects of New Zealand's community can gain access and be proficient in IT included the Cyber Communities Strategy, which have worked to give IT skills to the long-term unemployed.

As far as the notion of a digital divide in New Zealand society goes, one argument is that the divide is only present in the understanding, awareness and interest in Information Technology. On the other hand, IT can also be used to close traditional divides in New Zealand society. Although many lower socio-economic households do not have ownership of computers or high speed Internet, this does not mean they do not have access to information technology. These segments of New Zealand society may not have high speed Internet access at home; however through school, university, public libraries, and various community IT resources, they are able to access the Internet and gain vital IT skills needed in the workplace. This way it may be possible to build confidence and capability for all social levels of New Zealand Society. This argument presumes a widespread penetration of IT within New Zealand society and challenges the traditional notion of the digital divide by arguing that the divide is only present in the understanding of the capabilities and potential that IT can provide for individuals within NZ society.

The presentation *Diverse Dimensions of the 'Digital Divide'* from the Institute for the Study of Competition and Regulation in New Zealand took place at the Digital Divide in Asia-Pacific session at the Keio University program conference on designing governance for civil society in Tokyo on February 5th 2012. It provides perspectives on the digital divide in New Zealand and was delivered by Bronwyn Howell [6]. Its conclusions are drawn from academic research with an economic perspective. It provides a summary of the New Zealand governments policy responses to the so-called digital divide, discussing the unique circumstances with which New Zealand finds itself in an isolated part of the world but being highly developed and urbanised. The potential divides are highlighted as between rural and urban, densely populated areas and sparsely populated areas, as well as socioeconomic and demographic divides. Age, gender, ethnicity, income, and disability are possible areas of social divide in the digital era in New Zealand.

Other policy responses by the New Zealand government are summarised in this presentation such as the subsidised broadband network and rural broadband initiative. The government-subsidised ultra-fast broadband network aims to ensure 100mbps to 70% of New Zealand customers by 2018, prioritising schools and businesses to be upgraded first [6]. As mentioned previously, the rural broadband initiative is in partnership with Vodafone and Chorus to extend broadband in rural areas in order for 84% of rural customers to have access to broadband by 2016.

The conclusions drawn from the research report behind this presentation are that differences in prices and statistics are not necessarily evidence of a real social divide; and the assumption that rural consumers place a higher value on Internet connection and speed is not necessarily correct. The report calls for policy makers to understand underlying dimensions of a perceived divide before imposing polices that could in fact create further divisions. It advises governments to make well thought out, informed decisions when creating policies because ill-informed policies based on inaccurate research in this topic can prove costly. Pure statistics and Internet connection prices may paint an incomplete picture of a digital divide. Looking at different demographic and regional characteristics helps to provide a current day synopsis of the saturation of IT in New Zealand.

Voxy.co.nz featured a recent article, contributed by Fuseworks Media [4], about IT Minister Amy Adams attending a cyber policy conference. Voxy.co.nz is a New Zealand news website which also relies heavily on bloggers and real time news management to deliver current and relevant national news. During the conference in Budapest, Information technology Minister Amy Adams stressed that 2000 New Zealanders are affected by cyber crime every day and that this is a real issue that must be addressed by the government. Ms Adams stressed that cyber intrusions have the potential to disrupt our infrastructure, government and economy and that the conference was a chance for her to meet her counterparts and adopt new approaches to national cyber security. IT is integral to the political and social lives of our nation; in current times cyber security is an issue that every New Zealand computer user must address, and needs to be taken seriously by the government.

Stanford L Levin's paper on the Issues and Policies for Universal Service and Net Neutrality in a Broadband Environment focuses specifically on New Zealand as a basis for analysis and improvement. It was presented in Wellington (New Zealand) in August of 2010, and contains sections that discuss the situation in New Zealand from an interesting perspective. Due to the importance of broadband Internet for New Zealand society, it is incorporated in any universal service obligation. Australia and New Zealand both are undertaking government funded broadband network projects. On the other hand, New Zealand is one of the few exceptions of developed western countries where most broadband Internet packages have monthly caps. This supports the argument that there is still some way to go until New Zealand reaches the same standards as some other countries.

Levin's recommendations for New Zealand are that the country has unique characteristics and that the issue of universal service in this country is more centred on trying to achieve economies of scale rather than an ideological debate. Universal service is the provision of a basic standard to every resident in a country; in regards to Internet access, this has not yet been achieved in New Zealand.

Interestingly Levin argues that New Zealand's size is not an excuse to support an individual Internet Service Provider's monopoly. More competition should be encouraged although Telecom New Zealand currently has over the majority of the infrastructure. Competition in the provision of facilities and services is still possible, particularly with fibre and mobile services [9]. The issues with rural Internet provision are not unique to New Zealand and are faced most governments. However, New Zealand needs to constantly review and formulate its progressive Internet policies as it competes with other countries.

Erturk [3] discusses the international economics of information technology, specifically how information technology contributes to economic productivity and economic development in general. Erturk discusses the Internet as one of the great historical technology breakthroughs that has had a major impact on economies. It states that Internet usage is a sign of prosperity and the digital divide has been a source of concern within countries but also on an international level. The thesis goes on to discuss the importance of information technology in education and its effect on socio-economic development. It helps create human capital, which is considered to be a major economic resource. The countries which have grown fastest since the 1960's have invested heavily in human capital more so than those that have fallen relatively behind. This is an example of the affect information technology has on a country's economy, and the lessons can be applied to New Zealand. Computer technology can be used effectively in schools, libraries and throughout education to help build greater human capital [3]. Human capital is a major source of social-economic development, thus the use of IT in society and education will continue to have a long lasting effect on New Zealand society. By ensuring all New Zealanders have basic skills in information technology and access to high speed Internet, the quality of human capital in New Zealand society will continue to increase and, in turn, will aid growth and development.

Recently on Stuff.Co.NZ, Claire Rogers [14] summarises the latest findings by Statistics New Zealand in their recent study on broadband and mobile Internet use in New Zealand. Over half of New Zealand's population is currently using mobile Internet (that is 2.5 million people), with home broadband use also increasing massively in recent years. Home broadband users have increased more than 11% to 1.6 million customers over a year as Internet service providers offer more flexible packages with higher data caps. The number of users with 50gb or greater data caps has increased by 800%. 96% of subscribers now download at speeds of 1.5 to 24 Mbps [14].

Not only is Internet usage more widespread than ever before, but Internet speeds, data transfer and mobile Internet usage have also increased greatly. These findings are physical proof of the widespread use of information technology in New Zealand society and document the increasing spread and saturation of high speed Internet and mobile Internet technologies in New Zealand.

In *Connecting the Clouds,* Newman [11] details the history of telecommunications in New Zealand, particularly the Internet and how it has transformed government, business, communities and social lives. Keith Newman is the former news editor of *Computerworld*, editor of *Network World* and founding editor of *PC Magazine New Zealand*, and is a respected journalist who conducted over 100 interviews with scientists, computer programmers, telecommunications experts, engineers, business leaders and politicians in order to create an accurate history of IT in New Zealand. Three main chapters of the book are especially useful for analysing the impact of IT in New Zealand society.

The chapter entitled *"Clicks and Mortar, Beyond On-line Pamphlets"* [11] details the growth of e-commerce and business in New Zealand. Newman discusses how the NZ business environment began to change with the introduction of bar code readers, EFTPOS systems (electronic funds transfer at point of sale), and ATM's (automated teller machines). New Zealand has been one of the earliest countries in the adoption of barcode scanning and EFTPOS payments. For example, EFTPOS systems have been running all over New Zealand since the 1980's, and most retail transactions now rely on these systems. New Zealand society took to e-commerce and online retailing with vigour; Internet banking and online grocery shopping have been very successful in recent years and played a major role in changing New Zealand society. New Zealand's own online marketplace TradeMe is a success story. It was launched in 1999 by Sam Morgan and became very popular soon after. TradeMe initially specialised in computer parts but quickly branched out into antiques, clothing, cell phones and property. By October 2004, it was the most visited shopping site in New Zealand [11], Ebay is the global leader in online marketplace commerce however TradeMe was able to secure the New Zealand market by providing a local alternative that New Zealanders trust. According to the site's statistics, TradeMe has more than three million active members and approximately two million listings [16]. These numbers clearly demonstrate the popularity of TradeMe but also the willingness of New Zealanders to get involved in e-commerce and online retailing. This shows how quickly New Zealand society adapted to online e-commerce and ways in which the Internet could change everyday processes.

From a global perspective, in the 1990s, New Zealand lagged slightly behind similar sized countries (Finland, Israel, Singapore) in terms of the size of its Information Technology sector and its IT exports [2]. This was despite the facts that New Zealand enjoyed similar living standards otherwise, that the NZ government offered a large amount of funding for research and development, and that domestic IT use within New Zealand was as high as in the other countries. This may be attributed to the more distant location of New Zealand on the globe, and less capital available (especially private) for supporting and growing IT ventures in NZ, including both domestic and foreign investment. Competition on a global scale and integration with the rest of the world economy is one of the reasons why IT plays an important role in the public and government agenda in New Zealand.

There are currently a number of successful New Zealand software companies that have had a significant impact around the world, especially in other English-speaking countries such as the United States, United Kingdom, and Australia. These products, among many examples, include the JADE programming and software development platform, and more recently, Xero, which is an online accounting software package. Xero, founded in 2006 and expanded quickly since then with funds from American and Australian investors, works according to the cloud computing and software as a service model. It prides itself on being easy to use and easy to maintain, relatively to other major accounting software packages. It is available anywhere with Internet access, and on any mobile device that is connected to the Internet. Xero is also very popular locally in New Zealand due to many businesses willing to utilize available IT resources and keep up with the latest trends.

According to the Ministry of Economic Development [10], 96% of all NZ businesses use the Internet. Even among small and medium enterprises (i.e. business enterprises with less than 20 employees), the rate of Internet use (as part of their operations) is 88%. This Internet use includes banking, invoicing, and paying bills, among many other things.

Chapter Twenty entitled *"Digital Refresh Required, Government Learns to Share"* [11] documents the growth of e-government in New Zealand both in policy terms and in the ways with which New Zealanders interact with their government.  The government's IT policies and funding efforts have attempted to keep New Zealand up-to date and progressive in terms of digital strategy. New Zealand governments have traditionally put emphasis on IT growth and placed an importance on high technology use in health and education in order to further society as a whole. Information technology issues have, from time to time, dominated government policy and planning debates in New Zealand and will continue to do so.

Chapter Twelve, *"E-govt lumbers on-line, presenting a public face"* [11] analyses the specific uses of e-government in New Zealand. New Zealand was the first country in the world to launch an official government web server. Government websites are used by a majority of the population, for example, for form filling, registering cars, tax returns, and Inland Revenue processes online. It is clear that New Zealand has adapted to e-government very easily. According to a government survey, roughly half of New Zealand used the Internet to access government pages in 2008 [11]. http://www.govt.nz is the one stop portal to search and access all government forms and information from compiling a tax return to renewing a passport. It represents a push toward from the New Zealand government to create an e-democracy, which will involve instant public access to government information anytime, anywhere, and from any device [11]. The next challenge lies with ensuring all New Zealanders regardless of social background are proficient at doing these government processes online.

In summary, understanding the history of e-commerce and e-government in New Zealand helps provide a complete picture of IT development and Internet in New Zealand, as well as recent policies and their social impact.

## 3. Conclusion

It is clear that New Zealand society has been influenced greatly by developments in information technology. Policy makers and businesses must put great effort into understanding society's needs when developing current IT strategies. The emergence of 'Bring Your Own Device' computing in the workplace and the benefits of mobile technology in industry are being embraced by New Zealand businesses in order to remain competitive. This emerging trend is the future in global commerce and can greatly enhance business processes. By focusing on information technology in education and throughout the community New Zealand will be able to increase its human capital, which will in turn continue to support the country's overall growth and its international status. Mobile Internet in New Zealand is growing tremendously with more than half of the population; most New Zealanders are involved with online services and have a basic grounding in information technology. The success of online tools such as TradeMe, Xero, and the NZ government's online portal demonstrates that New Zealand is willing to quickly embrace new and emerging online activities and moving towards becoming a true 'e-society.' New Zealand is an IT driven society and will continue to develop in this area in the years to come.

The review of existing literature makes it clear that household income and Internet use are linked. There are clear variations in income, Internet usage and other statistics in NZ, all of which may imply a vast digital divide between lower socio-economic groups and the urbanised middle classes. Although it is being addressed through government policy, there is somewhat of a digital divide between rural and urban New Zealand, which is also a global phenomena. Schools, libraries, universities and community centres throughout New Zealand make it possible for all segments of society to have access to the Internet and gain general skills in information technology. The continued focus by the government through IT initiatives such as the Digital Opportunities and Rural Broadband Initiative aim at reducing this divide and ensuring all New Zealanders have Internet access and a basic grounding in IT. It therefore clear that the risk of such a divide is understood by the NZ government. While working on this goal, the NZ government must be careful not limit its efforts to short-term fixes or occasionally giving away free goods to disadvantaged groups as past studies on domestic digital divide in other countries such as the USA suggest that the divide will persist unless the underlying economic and educational gaps also change [7].

Current policy is geared around providing a universal service to all New Zealanders. Levin [9], discussing net neutrality and universal service, provided a sound argument that the size of New Zealand is not an excuse for the current monopoly held over infrastructure by Telecom New Zealand, and competition should be encouraged without using the economies of scale as an counter argument. Although New Zealand has come a long way, there is room for improvement. More competition between Internet Service Providers should be encouraged. The fact that the majority of Internet packages still have capped data allowances is a somewhat backward situation.

The potential digital divide between lower and upper income households, and between rural and urban areas, is being addressed by ongoing government IT initiatives and polices. The existence of any disadvantaged individual households (due to income, etc.) may be addressed through social policy and initiatives. All New Zealanders are provided with a basic knowledge of IT in school, while the ability to interact with technology and the Internet regularly can still be enhanced through education and community projects. Household income will always limit some parts of the community with regards to their ability to stay at the forefront of emerging technologies and skills but, through education and community funding, the negative impact of this divide can be limited.

Although the provision and access to IT is available to every segment of New Zealand society, the desire to learn to use IT along with understanding the importance and benefits of IT in the workplace is not universally shared throughout society. Without the motivation to develop new skills and knowledge, lower socio-economic groups will be left disadvantaged as society and the working environment continues to develop and progress into an increasingly technological world. Education is the key to ensuring this divide does not increase and spread, if all aspects of society are to be made aware of the importance of IT and the opportunities and funding are available to everyone regardless of income or location, so that New Zealand will continue to develop progressively.

Using this research as a basis it is advisable to conduct further studies within relatively lower socio-economic areas of New Zealand in order to investigate whether or not the communities in the area are aware of all of the opportunities available to them with regards to Information Technology and to understand these individuals' views on the importance of IT in modern New Zealand. One such study was done recently but outside of New Zealand. Focus groups were held with 80 disadvantaged and low-income Australians to have a closer analysis of factors that lead to domestic digital divide, and the reasons why these individuals utilize Internet and digital resources, and do not benefit equally from online public and health services [12]. The study identified some barriers that affect disadvantaged individuals such as: technological literacy, education, income, housing situation, social connection, health status, and employment status.

A quantitative survey and qualitative interviews with high school students in areas identified as low income and in low decile school zones, may provide an accurate insight into these communities' awareness and understanding of information technology and their own satisfaction with current government initiatives and policies. A future study in some of the rural areas of New Zealand may provide slightly different but equally valuable insights. With the agriculture industry being the most dominant in New Zealand, a study into the development of information technology in rural communities would prove useful in assessing the success of the rural broadband initiative and provide quantitative results with which to assess any potential digital divide. Another area that can be pursued is the marginalisation of the older workforce whose past experience and training was done prior to IT being widely introduced into the work place. Although a significant percentage of older people have sought to keep themselves aware and active in the use of new technology, it is likely that a majority of them still struggles with IT in their lives whereby they are impeded, restricted or isolated because of a lack of certain skills, or a lack of confidence when they are required to use IT for with government and social services access.

**References**

1. Cruz-Jesus, F., Oliveira, T., & Bacao, F. (2012). Digital divide across the European Union. *Information & Management*.

2. Ein-Dor, P., Myers, M. and Raman, K.S., (2004), IT Industry Development and the Knowledge Economy: A Four Country Study, *Journal of Global Information Management*, 12, 4, 23-49.

3. Erturk, E. (2007) *Studies on the international economics of information technology*. (Doctoral Thesis). Proquest Central. UMI No. 3257949.

4. *Fuseworks Media* (2012, October 2). *Minister to Attend Cyber Policy Conference*. Retrieved from: http://www.voxy.co.nz/politics/minister-attend-cyber-policy-conference/5/136511

5. Head, M. (2004). *Addressing the 'Digital Divide' in Aotearoa New Zealand*. The Finnish Encyclopaedia of Digital Government. Retrieved from: http://www.uta.fi/laitokset/ISI/EnDigG/

6   Howell, B. (2012, February 5). *Diverse Dimensions of the 'Digital Divide': Perspectives From New Zealand.* Presented at the session Digital Divide in Asia-Pacific at the Keio University Global COE Programme Conference on Designing Governance for Civil Society, Tokyo. Retrieved from: http://www.iscr.org.nz/f718,19865/Dimensions_of_Digital_Divide.pdf

7   Kvasny, L. and Keil, M. (2006), The challenges of redressing the digital divide: a tale of two US cities. Information Systems Journal, 16: 23–53.

8   Kevany, K. (2010, August). The new age of business mobile. *NZ Business Magazine.* Retrieved from http://nzbusiness.co.nz/articles/new-age-mobile

9   Levin, S. (2010,September). *Issues and Policies for Universal Service and Net Neutrality in a Broadband Environment*. The Institute for the Study of Competition and Regulation Inc. Retrieved from: http://www.iscr.org.nz/f605,17350/17350_Universal_Service_and_Net_Neutrality_in_Broadband_-_Final.pdf

10  Ministry of Economic Development (2011). *SME's in New Zealand: structure and dynamics.* Retrieved from: http://www.med.govt.nz/business/business-growth-internationalisation/pdf-docs-library/structure-and-dynamics-2011.pdf

11  Newman,K. (2008). *Connecting the Clouds, The Internet in New Zealand*. Auckland, New Zealand. Activity Press

12  Newman, L., Biedrzycki, K., & Baum, F. (2012). Digital technology use among disadvantaged Australians: implications for equitable consumer participation in digitally-mediated communication and information exchange with health services. *Australian Health Review*, *36*(2), 125-129.

13  Norris, P. 2001. Digital divide: Civic engagement, information poverty and the Internet world-wide. Cambridge, MA: Cambridge Univ. Press.

14  Rodgers,C. (2012, October 12) *2.5 Million Kiwis using Mobile Internet*. Retrieved from: http://www.stuff.co.nz/technology/digital-living/7808200/2-5-million-Kiwis-using-mobile-Internet

15  Statistics New Zealand. (2004). The Digital Divide- *Examining the main characteristics that influence household Internet Connection in New Zealand*. Retrieved from www.stats.govt.nz http://www.stats.govt.nz/browse_for_stats/industry_sectors/information_technology_and_communications/digital-divide.aspx

16  TradeMe. (2012). *Site Stats September 2012*. Retrieved from: http://www.trademe.co.nz/About-trade-me/Site-stats

17  Willis, D. (2012, August). *Bring your own device: New opportunities, New challenges*. www.gartner.com. Retrieved from: http://www.gartner.com/DisplayDocument?doc_cd=238131&ref=g_noreg

**AUTHOR PROFILES**

1. Emre Erturk received his Ph.D. from the University of Oklahoma in 2007. He has later taught as an Assistant Professor with the University of Maryland. Currently, he is a Senior IT Lecturer in the School of Computing at the Eastern Institute of Technology in New Zealand. He has been involved in many conferences and publications around the world.

2. Derwyn Fail received his Bachelor's degree with Honours in History from Swansea University in Wales (UK). Next, he completed his Graduate Diploma in Information Technology at the Eastern Institute of Technology (EIT) in New Zealand in 2012. He has been a research student of Dr. Emre Erturk, and started this paper both as a research project and to inform his future career in public service.